# Predictable Migration and Communication in the Quest-V Multikernel

Ye Li, Eric Missimer and Richard West

*Computer Science Department*
*Boston University*
*Boston, MA 02215*
*Email: {liye,missimer,richwest}@cs.bu.edu*

*Abstract*—Quest-V is a system we have been developing from the ground up, with objectives focusing on safety, predictability and efficiency. It is designed to work on emerging multicore processors with hardware virtualization support. Quest-V is implemented as a "distributed system on a chip" and comprises multiple sandbox kernels. Sandbox kernels are isolated from one another in separate regions of physical memory, having access to a subset of processing cores and I/O devices. This partitioning prevents system failures in one sandbox affecting the operation of other sandboxes. Shared memory channels managed by system monitors enable inter-sandbox communication.

The distributed nature of Quest-V means each sandbox has a separate physical clock, with all event timings being managed by per-core local timers. Each sandbox is responsible for its own scheduling and I/O management, without requiring intervention of a hypervisor. In this paper, we formulate bounds on inter-sandbox communication in the absence of a global scheduler or global system clock. We also describe how address space migration between sandboxes can be guaranteed without violating service constraints. Experimental results on a working system show the conditions under which Quest-V performs real-time communication and migration.

## I. INTRODUCTION

Quest-V is a real-time system we are developing that focuses on predictability, efficiency and safety. It is built from the ground up rather than being a modification to a pre-existing system such as Linux. We chose this path because Linux is in many ways overly complex, making it difficult to enforce system-wide predictability by simply modifying individual components such as the scheduler. Additionally, our development of Quest-V has enabled us to investigate ways to integrate emerging hardware features directly into the design of the OS. For example, many modern processors feature hardware virtualization support (e.g., Intel VT, AMD-V and ARM Cortex A15 processors), and Quest-V uses these capabilities to isolate system components into separate "sandbox kernels". This enhances fault tolerance and dependability of the system. Unlike traditional hypervisors that support separate largely unrelated guest OSes and their applications, Quest-V sandbox kernels work together as a distributed system on a chip.

The distributed organization of Quest-V means there is no notion of a single kernel responsible for global scheduling or system-wide resource management. Instead, each sandbox kernel manages its own dedicated resources, including a specific set of cores, I/O devices and memory regions. The philosophy behind Quest-V is to share as little as possible between different sandbox kernels and their memory spaces. This reduces resource contention and has been shown to increase scalability [1]. However, this imposes several challenges on the design of applications. For example, multi-threaded applications must be distributed across the various sandbox kernels to increase the degree of parallelism. Since each sandbox manages a private region of physical memory, separate threads in different sandboxes will each need a private copy of their address spaces. Parallel applications must be designed to work in a distributed manner, communicating across special shared memory channels when necessary.

Quest-V's distributed system design means that each sandbox operates with an independent system clock. Just as traditional distributed systems have neither physically shared memory nor a single physical clock for all loosely coupled compute nodes, each Quest-V sandbox on a multicore processor uses a local timer on each core to manage event timings.

The lack of a global clock and a global scheduler in Quest-V poses challenges in two key ways: (1) in the communication between threads in different sandboxes, and (2) the migration of threads between sandboxes. The first problem occurs when two or more threads need to synchronize or exchange data within time bounds. A sending thread in one sandbox may be forced to wait for a reply from a remote thread that is independently scheduled in another sandbox. A method to bound the round-trip delay is necessary. The second problem occurs when a thread may have partially executed in one sandbox and needs to complete by a certain time in another

This work is supported in part by NSF Grant #1117025.

sandbox. Ideally, we want the migrating thread to avoid any penalty for migration and clock skew that could affect its ability to meet timing requirements. Likewise, any migrating thread should not adversely affect the timing guarantees of threads already executing in the destination sandbox.

**Contributions.** This paper describes how Quest-V guarantees predictable inter-sandbox communication and thread migration between separate sandboxes running on a multicore processor. We formulate constraints under which these objectives can be guaranteed without requiring a common clock or global scheduler.

In the next section, we briefly describe the Quest-V architecture. This is followed by a discussion of the mechanisms to enforce predictable thread migration, as well as inter-sandbox communication. Experimental results are shown in Section V. A summary of related work is described in Section VI. Finally, conclusions and future work are discussed in Section VII.

## II. QUEST-V ARCHITECTURE

As stated earlier, the Quest-V system is partitioned into a series of *sandbox kernels*, with each sandbox encompassing a subset of memory, I/O and CPU resources. Sandbox memories are isolated from one another using hardware virtualization capabilities found on modern multicore processors. The current system works on x86 (Intel VT-x) platforms but plans are underway to port Quest-V to the ARM architecture.

We envision Quest-V being applicable to safety-critical systems, where it is important that services remain operational even when there are hardware and software failures. Future automotive, avionics, factory automation and healthcare systems are example application domains where safety-criticality is required.

Although we assume the existence of hardware virtualization features, this is not essential for the design of our system. Sandbox memory isolation can instead be enforced using segmentation or paging on platforms with simpler memory protection units. However, hardware virtualization offers potential advantages (albeit with added costs) that we wish to investigate in our system design. In particular, it provides an extra logical "ring of protection" that allows services to execute with traditional kernel privileges. This is in contrast with micro-kernels that grant the least privileges necessary for services, at the expense of added communication to access other services in separate protection domains.

A high-level overview of the Quest-V architecture is shown in Figure 1. Each sandbox, including its kernel and application space, is associated with a separate monitor. The monitors are relatively small, fitting easily into a $4KB$ memory page. They primarily exist to maintain extended page table (EPT) mappings, so that each sandbox's virtual memory space maps onto a separate region of host physical memory. The only other times monitors are needed is to launch a sandbox when the system is initialized, to aid in fault recovery or protection management (e.g., if there is an EPT violation), and to update shared memory mappings between sandboxes.

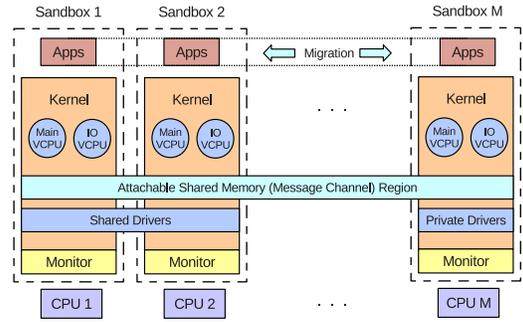

Fig. 1. Quest-V Architecture Overview

Shared memory mappings are used in Quest-V to establish communication channels between the otherwise separate sandboxes. We use replicated monitors rather than a single hypervisor such as in Xen [2] for three main reasons. First, each monitor is a trusted code base with a small memory footprint (less than 4KB). Second, the monitors can be implemented differently so that they are not all susceptible to the same security vulnerability. Third, a monitor maintains EPT mappings for only one sandbox, eliminating the overheads of scheduling and switching guest address spaces as is done with traditional hypervisor systems.

In Quest-V all scheduling and device management is performed within each sandbox directly. We use a form of I/O passthrough to allow device interrupts to be directed to a sandbox kernel without monitor intervention. This differs from the "split driver" model of systems such as Xen that have a special domain to handle interrupts before they are directed into a guest. Allowing sandboxes to have direct access to I/O devices and to perform local scheduling decisions greatly reduces the overhead of switching into a hypervisor (or, equivalently, monitor) to aid in resource management.

Quest-V supports configurable partitioning of resources amongst guests, similar to the way cores are partitioned in systems such as Corey [3]. By default, we assume each sandbox is associated with a single processing core since this simplifies local (sandbox) scheduling, although it is possible to configure a sandbox to encompass multiple cores. Similarly devices can be shared or partitioned. For example, a network device could be exclusive to one sandbox while all USB devices are shared between two or more sandboxes.

An IO APIC is programmed to deliver interrupts from a device to all sandboxes with access to that device. Device drivers are written so they perform *early demultiplexing* [4] of device interrupts, discarding any interrupts that are not ultimately associated with a device request in the local sandbox. While this is out of the scope of this paper, we have shown this method of device sharing and interrupt delivery results in lower latency and higher throughput than using a split driver as in Xen [5].

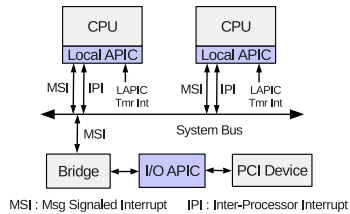

Fig. 2. APIC Configuration

Figure 2 shows the hardware APIC configuration. Each Local APIC, associated with a different core, is used to generate IPIs and establish local timer events.

### A. VCPU Scheduling and Migration

Quest-V uses virtual CPUs (VCPUs) as the basis for time management and predictability of its sub-systems. VCPUs in Quest-V differ from those in conventional virtualized systems. Rather than maintaining guest machine state, they act as resource containers for shares of actual CPU time, and are assigned priorities for scheduling purposes. Multiple threads can share the same VCPU, with the chosen thread being assigned according to some configurable policy (e.g., FCFS, rate-monotic, EDF, or a priority inheritance-based approach). VCPUs are then scheduled on available physical CPUs (PCPUs) by the local sandbox kernel.

By default, VCPUs act like Sporadic Servers [6], and are assigned static priorities. Each VCPU, $V_i$, has a budget capacity, $C_i$, and replenishment period, $T_i$. Rate monotonic scheduling [7] can then be applied to determine schedulability. For improved utilization it is possible to configure Quest-V to schedule VCPUs in increasing deadline order. However, for cases where there are multiple tasks with different deadlines sharing the same VCPU, there is an increased overhead of managing VCPU priorities dynamically. Additionally, with real-time multicore systems, we see predictable resource management as being more important than maximizing total CPU utilization in all cases. Factors such as shared caches and other micro-architectural resource contention [8], [9] affect thread progress on multicore systems, and are arguably as important as CPU utilization.

Quest-V decouples the scheduling of conventional tasks from those associated with I/O events, such as interrupts. The latter class of events is often sporadic in nature and may be associated with a thread that originated an I/O request. Quest-V integrates interrupt and task scheduling, preventing interrupts from arbitrarily interfering with task execution. While conventional tasks execute on Main VCPUs, their I/O events are processed in the context of I/O VCPUs (See Figure 3). Every I/O event is associated with an I/O VCPU, whose priority is inherited from a corresponding Main VCPU. Thus, a thread, $\tau$, running on Main VCPU, $V_M$, may block while awaiting the response from a service request handled by an I/O VCPU, $V_{IO}$. The priority of $V_{IO}$ is inherited from that of $\tau$'s Main VCPU, after which $\tau$ can be woken up to resume on $V_M$. In past work, we have shown how this integrated interrupt and task scheduling approach is more predictable than systems such as Linux [10].

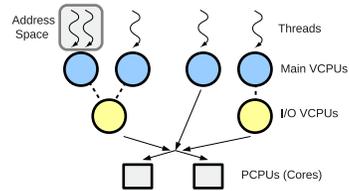

Fig. 3. VCPU Scheduling Hierarchy

In Quest-V there is no notion of a periodic timer interrupt for updating system clock time. Instead, the system is event driven, using per-processing core local APIC timers to replenish VCPU budgets as they are consumed during thread execution.

### B. Inter-Sandbox Communication

Inter-sandbox communication in Quest-V relies on message passing primitives built on shared memory, and asynchronous event notification mechanisms using Inter-processor Interrupts (IPIs). IPIs are currently used to communicate with remote sandboxes to assist in fault recovery, and can also be used to notify the arrival of messages exchanged via shared memory channels. Monitors update extended page table mappings as necessary to establish message passing channels between specific sandboxes. Only those sandboxes with mapped shared pages are able to communicate with one another. All other sandboxes are isolated from these memory regions.

A *mailbox* data structure is set up within shared memory by each end of a communication channel. By default, Quest-V currently supports asynchronous communication by polling a status bit in each relevant mailbox to determine message arrival. Message passing threads are bound to VCPUs with specific parameters to control the rate of exchange of information. Likewise, sending and receiving threads are assigned to higher priority VCPUs to reduce the latency of transfer of information across a communication channel. This way, shared memory channels can be prioritized and granted higher or lower throughput as needed, while ensuring information is communicated in a predictable manner. Thus, Quest-V supports real-time communication between sandboxes without compromising the CPU shares allocated to non-communicating tasks.

The lack of both a global clock and global scheduler for all sandboxes creates challenges for a system requiring strict timing guarantees. In the next two sections we elaborate on two such challenges, relating to predictable address space migration and communication.

### III. PREDICTABLE MIGRATION FRAMEWORK

Quest-V allows threads and VCPUs to be migrated between sandboxes, rather than forcing them to be statically mapped. There are several reasons why migration might be desired, including:

*a) performance:* by redistributing workloads across sandboxes we can avoid pathological co-schedules that conflict with micro-architectural resources. Such resources include shared on-chip caches and memory buses. Without judicious co-scheduling, it is possible for thread progress to be stalled even if VCPU shares are guaranteed.

*b) predictability:* dynamically-created threads and VCPUs can overload the local sandbox, making it impossible to guarantee their timing requirements without migration to less loaded sandboxes.

*c) resource affinity:* a thread may require access to resources, such as I/O devices, that are not available in the local sandbox. Either the thread can communicate via shared memory to a remote sandbox, potentially incurring high latencies, or it can be migrated to the sandbox that owns the required resources.

*d) fault recovery:* a software fault in the local sandbox might render a service required by a thread inoperable. As part of system recovery, the local service can be restored in the background while the thread continues in a different sandbox.

For this paper, we do not focus on policies to decide where a thread should be migrated or how to recover from faults. Rather, given that migration is supported, the key issue is how to guarantee a thread and its VCPU's timing guarantees when it is migrated.

#### A. Predictable Migration Strategy

Threads in Quest-V have corresponding address spaces and VCPUs. The current design limits one, possibly multi-threaded, address space to be associated with a single VCPU. This restriction avoids the problem of migrating VCPUs and multiple address spaces between sandboxes, which could lead to arbitrary delays in copying memory. Additionally, only Main VCPUs and their address spaces are migratable, as I/O VCPUs are pinned to sandboxes with access to specific devices.

Migration from one sandbox's private memory requires a copy of an address space and all thread data structures to the destination. Each thread is associated with a `quest_tss` structure that stores the execution context and VCPU state.

Figure 4 shows the general migration strategy. Although not shown, both application address spaces and kernel threads are migratable. An inter-processor interrupt (IPI) is first sent to the destination sandbox, to initiate migration. A special *migration thread* handles the IPI in the destination, generating a trap into its monitor that has access to machine physical memory of all sandboxes. The migrating address space in the source sandbox is temporarily mapped into the destination. The address space and associated `quest_tss` thread structures are then copied to the target sandbox's memory. At this point, the page mappings in the source sandbox can be removed by the destination monitor. To bound the costs of migration, a limit is placed on the number of threads and, hence, `quest_tss` structures, within a single address space.

An IPI from the destination to the source sandbox is needed to signal the completion of migration. This is handled by a migration thread in the source sandbox, which is able to reclaim the memory of the migrated address space. All IPIs are handled in the sandbox kernels, with interrupts disabled while in monitor mode. The migration thread in the destination can now exit its monitor and return to the sandbox kernel. The migrated address space is attached to its VCPU and added to the local schedule. At this point, the migration threads in source and destination sandboxes are able to yield execution to other VCPUs and, hence, threads.

Before migrating an address space and its VCPU, the migration thread in the destination sandbox performs admission control. This verifies the scheduling requirements of the incoming VCPU can be guaranteed without violating the requirements of all existing VCPUs.

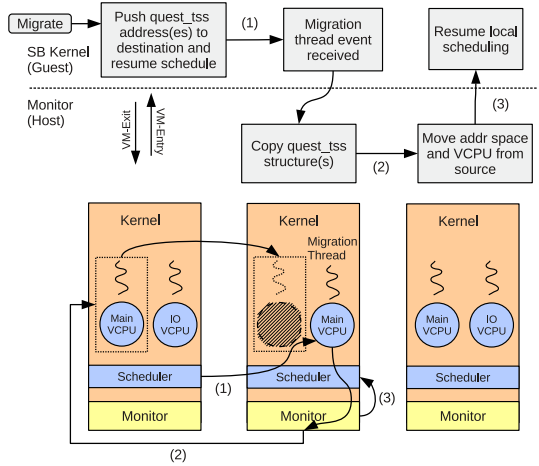

Fig. 4. Migration Strategy

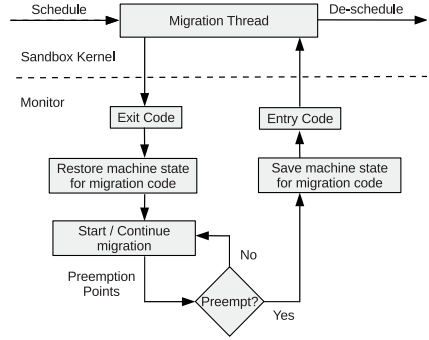

Fig. 5. Migration Framework Control Flow

**Migration Thread Preemption.** A migration thread in the destination sandbox is bound to a Main VCPU with parameters $C_m$ and $T_m$. If the VCPU depletes its budget, $C_m$, or a higher priority VCPU is ready to run, the migration thread should be preempted. However, this is complicated by the fact that the migration thread spends most of its time inside its monitor (a.k.a. VMX root mode) and each sandbox scheduler runs within the local kernel (a.k.a. VMX non-root mode).

Migration thread preemptions require a domain switch between a sandbox monitor and its kernel, to access the local scheduler. This results in costly VM-Exit and VM-Entry operations that flush the TLB of the processor core. To avoid this cost, we limited migration thread preemption to specific preemption points. Additionally, we associated each migration thread with a highest priority VCPU, ensuring it would run until either migration was completed or the VCPU budget expired. Bookkeeping is limited to tracking budget usage at each preemption point. Thus, within one period, $T_m$, a migration thread needs only one VM-Exit and one VM-Entry.

Preemption points are currently located: (1) immediately after copying each `quest_tss` structure, (2) between processing each Page Directory Entry during address space cloning, and (3) right before binding the migrated address space to its VCPU, for re-scheduling. In the case of a budget overrun, the next budget replenishment is adjusted according to the corrected POSIX Sporadic Server algorithm [11]. Figure 5 describes the migration control flow.

**Clock Synchronization.** One extra challenge to be considered during migration is clock synchronization between different sandboxes in Quest-V. Quest-V schedulers use Local APIC Timers and Time Stamp Counters (TSCs) in each core as the source for all time-related activities in the system, and these are not guaranteed to be synchronized by hardware. Consequently, Quest-V adjusts time for each migrating address space to compensate for clock skew. This is necessary when updating budget replenishment and wakeup time events for a migrating VCPU that is sleeping on an I/O request, or which is not yet runnable.

The source sandbox places its current TSC value in shared memory immediately before sending a IPI migration request. This value is compared with the destination TSC when the IPI is received. A time-adjustment, $\delta_{ADJ}$, for the migrating VCPU is calculated as follows:

$$\delta_{ADJ} = TSC_d - TSC_s - 2 * RDTSC_{cost} - IPI_{cost}$$

$TSC_d$ and $TSC_s$ are the destination and source TSCs, while $RDTSC_{cost}$ and $IPI_{cost}$ are the average costs of reading a TSC and sending an IPI, respectively. $\delta_{ADJ}$ is then added to all future budget replenishment and wakeup time events for the migrating VCPU in the destination sandbox.

### B. Migration Criteria

Quest-V restricts migratable address spaces to those associated with VCPUs that either: (1) have currently expired budgets, or (2) are waiting in a sleep queue. In the former case, the VCPU is not runnable at its foreground priority until its next budget replenishment. In the latter case, a VCPU is blocked until a wakeup event occurs (e.g., due to an I/O request completion or a resource becoming available). Together, these two cases prevent migrating a VCPU when it is runnable, as the migration delay could impact the VCPU's utilization.

For VCPU, $V_s$, associated with a migrating address space, we define $E_s$ to be the *relative time* [1] of the next event, which is either a replenishment or wakeup. For the utilization of $V_s$ to be unaffected by migration, the following must hold:

$$E_s \geq \lfloor \frac{\Delta_s}{C_m} \rfloor \cdot T_m + \Delta_s \bmod C_m, \quad (1)$$

where $C_m$ and $T_m$ are the budget and period of the migrating thread's VCPU, and $\Delta_s$ is the migration cost of copying an address space and its *quest_tss* data structures to the destination. At boot time, Quest-V establishes base costs for copying memory pages without caches enabled [2]. These costs are used to determine $\Delta_s$ for a given address space size. Quest-V makes sure that the migrating thread will not be woken up by asynchronous events until the migration is finished. The system imposes the restriction that threads waiting on I/O events cannot be migrated.

---

[1] i.e., Relative to current time.

A schedulability test is performed before migration, to ensure a VCPU can be added to the destination sandbox without affecting total utilization. If the test fails, the migration request will be rejected immediately by an IPI, and the source sandbox will put the address space and its VCPU back into the local scheduler queue. At the next scheduling point, a new destination sandbox is tested if necessary. A VCPU can be migrated immediately for any successful test, if it does not require its utilization to be guaranteed while migration is in progress.

In order to simplify the migration criteria, our current implementation restricts concurrent migration requests to different destination sandboxes. This is not problematic as migrations are expected to be infrequent.

## IV. Predictable Communication

In Quest-V, as in any distributed system, there is often a need for threads to communicate and exchange information. This section describes how Quest-V attempts to guarantee bounded and predictable communication latency between threads in different sandboxes.

Consider a sending thread, $\tau_s$, associated with a VCPU, $V_s$, which wishes to communicate with a receiving thread, $\tau_d$, bound to $V_d$ in a remote sandbox. Suppose $\tau_s$ sends a message of $N$ bytes at a cost of $\delta_s$ time units per byte. Similarly, suppose $\tau_d$ replies with an $M$ byte message at a cost of $\delta_d$ time units per byte. Before replying, let $\tau_d$ consume $K$ units of processing time to service the communication request.

[2]We do not consider memory bus contention issues, which could make worst-case estimations even larger.

We first define the following symbols:

$$L_s = C_s - (N \cdot \delta_s) \bmod C_s$$
$$L_d = C_d - (M \cdot \delta_d + K) \bmod C_d$$
$$S = \lceil \frac{N \cdot \delta_s}{C_s} \rceil \cdot T_s - L_s$$
$$D = \lceil \frac{M \cdot \delta_d + K}{C_d} \rceil \cdot T_d - L_d$$
$$R = \frac{D - L_s}{T_s}$$
$$Q = D - L_s - \lfloor R \rfloor \cdot T_s$$
$$P = Q - (T_s - C_s)$$
$$B = C_s - f(\frac{Q}{T_s - C_s}) \cdot P$$

where

$$f(x) = \begin{cases} 1 & \text{if } x \geq 1 \\ 0 & \text{if } x < 1 \end{cases}$$

Assuming $\tau_s$ starts communicating with $\tau_d$ at the end of its current budget, the worst case round trip time ($W$) as shown in Figure 6 is:

$$W = S + D + f(\frac{D}{L_s}) \cdot (L_s + \lceil R \rceil \cdot T_s - B - D) \quad (2)$$

Equation 2 is derived from the 5 different scenarios, as shown in Figure 6, for when the receiver (Recv 1-5) responds to the sender. This is dependent on VCPU budget availabilities of the sender and receiver. In the equation, $S$ represents the total time taken by $\tau_s$ to send a request message. $D$ is the total time between the end of $\tau_s$'s request (i.e. end of $S$) and the end of $\tau_d$'s response. $L_s$ is the remaining budget of $V_s$ after sending a request. Similarly, $L_d$ is the remaining budget of $V_d$ after sending a response. $R$ is the number of elapsed periods of $V_s$ to receive a response to $\tau_s$'s request. $Q$ is the time between $\tau_d$'s response and the end of $\tau_s$'s most recently used budget. $P$ is the partially consumed current budget for $\tau_s$ when it receives a response. $P$ is 0 for cases when $\tau_d$ completes its response and $\tau_s$ is not runnable. Finally, $B$ is the remaining budget for $\tau_s$ when it receives a response. In essence, Equation 2 becomes $W = S + D$ for Case 1, and $S + L_s + \lceil R \rceil \cdot T_s - B$ for all other cases.

Figure 6 assumes $\frac{N \cdot \delta_s}{C_s} < 1$ and $\frac{M \cdot \delta_d + K}{C_d} < 1$. This means, a request is processed in the budget capacity, $C_s$, and a response is processed in $C_d$. $S$ and $D$ capture the general cases where processing time may exceed the full budget capacity of the sender or receiver.

The worst case round trip time shown in Equation 2 can be exceeded in certain cases where the sender starts communication before the end of its current budget. Figure 7 shows the situation where $\tau_s$ starts communicating at $S'$, while it still has some budget remaining. Consider case 4 as an example, if we shift $S$ to the left such that $\tau_s$ finishes at the beginning of its current budget, we can actually increase the worst case time by $C_s - L_s$. Hence the worst case round trip time ($W'$) becomes:

$$W' = S + D + f(\frac{D}{L_s}) \cdot (L_s + \lceil R \rceil \cdot T_s - B - D) + E \quad (3)$$

where $E$ is defined as:

$$E = max(N_1, N_2)$$
$$N_1 = (1 - f(\frac{Q}{T_s - C_s})) \cdot min(Q, C_s - L_s)$$
$$N_2 = f(\frac{Q}{T_s - C_s}) \cdot f(\frac{C_s - L_s}{P}) \cdot$$
$$min(T_s - C_s, C_s - L_s - P)$$

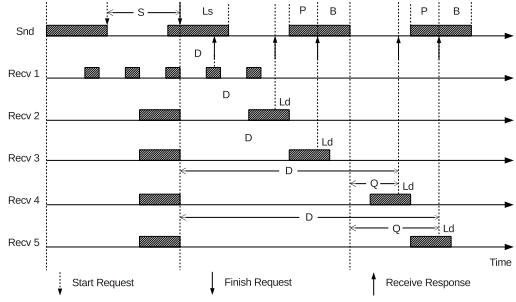

Fig. 6. Communication Cost Scenarios

Here, $E$ is the extra time added to Equation 2, caused by the shift of $S$ to $S'$. $N_1$ is the term representing cases 2 and 4, while $N_2$ covers cases 3 and 5.

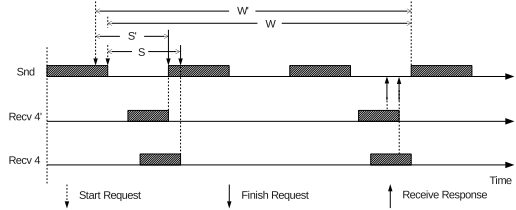

Fig. 7. Phase shift of S

## V. EXPERIMENTAL EVALUATION

We conducted a series of experiments on a Gigabyte Mini-ITX machine with an Intel Core i5-2500K 3.3GHz 4-core processor, 8GB RAM and a Realtek 8111e NIC.

### A. Predictable Migration

To verify the predictability of the Quest-V migration framework, we constructed a task group consisting of 2 communicating threads and another CPU-intensive thread running a Canny edge detection algorithm on a stream of video frames. The frames were gathered from a LogiTech QuickCam Pro9000 camera mounted on our RacerX mobile robot, which traversed one lap of Boston University's indoor running track at Agganis Arena [3]. To avoid variable bit rate frames affecting the results of our experiments, we applied Canny repeatedly to the frame shown in Figure 8 rather than a live stream of the track. This way, we could determine the effects of migration on a Canny thread by observing changes in processing rate while the other threads communicated with each other.

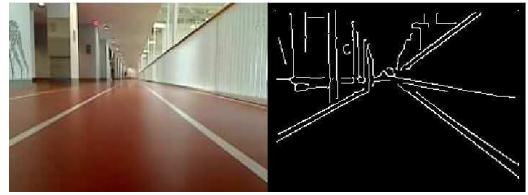

Fig. 8. Track Image Processed by Canny

For all the experiments in this section, we have two active sandbox kernels each with 5 VCPUs. The setup is shown in Table I. The Canny thread is the target for migration from sandbox 1 to sandbox 2 in all cases. Migration always starts at time 5. A logger thread was used to collect the result of the experiment in a predictable manner. Data points are sampled and reported in a one second interval.

| VCPU (C/T) | Sandbox 1 | Sandbox 2 |
|---|---|---|
| 20/100 | Shell | Shell |
| 10/50 | Migration Thread | Migration Thread |
| 20/100 | Canny | |
| 20/100 | Logger | Logger |
| 10/100 | Comms 1 | Comms 2 |

TABLE I
MIGRATION EXPERIMENT VCPU SETUP

Figure 9 shows the behavior of Canny as it is migrated in the presence of the two communicating threads. The left y-axis shows both Canny frame rate (in frames-per-second, *fps*) and message passing throughput (in multiples of a 1000 Kilobytes-per-second). The right y-axis shows the actual CPU consumption of the migration thread in (millions of, *x1m*) cycles. We can see from this figure that none of the threads (2 communicating threads and Canny) have been affected due to migration. The sudden spike in migration thread CPU consumption occurs during the migration of the Canny thread.

The average time to migrate an address space varies from under 1 millisecond to about 10 - 20 milliseconds, depending on the actual address space size. This is with all caches enabled and with address spaces being limited to a maximum of $4MB$. As long as the criteria in Section III-B are satisfied, all VCPUs remain unaffected in terms of their CPU utilization during migration.

---

[3]RacerX is a real-time robot control project that leverages Quest-V.

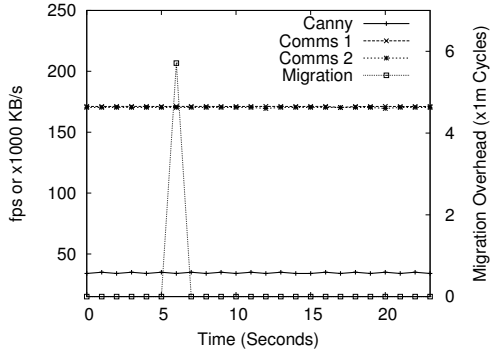

Fig. 9. Migration With No Added Overhead

Table II shows the values of variables as defined in Equation 1. The worst-case migration cost, $\Delta_s, worst$, was the cost of copying a Canny address space with all caches disabled (including the overhead of walking its page directory). $\Delta_s, actual$ was the actual migration thread budget consumption during migration. Both worst-case and actual migration costs satisfy the constraints of Equation 1.

| Variables | $E_s$ | $\Delta_s, worst$ | $\Delta_s, actual$ | $C_m$ | $T_m$ |
|---|---|---|---|---|---|
| Time (ms) | 79.8 | 5.4 | 1.7 | 10 | 50 |

TABLE II
MIGRATION CONDITION

The next experiment investigated the effect on the migrating address space when Equation 1 cannot be satisfied. This may impact the predictability of the migrating address space. If this is acceptable then the system will transfer the address space. We added a busy-wait overhead of $800\mu s$ to the address space clone procedure for each processed Page Directory Entry (of which there were 1024 in total). This increased migration costs in a controlled manner.

Figure 10 shows how the migration costs increase, with only the migrating address space being affected. This is observed by the drop in Canny frame processing rate, but all other threads and their VCPUs are unaffected. Here, the preemption points within each sandbox monitor prevent excessive budget overruns that would otherwise impact VCPU schedulability.

Table III shows the migration parameters for this experiment. The drop in Canny frame rate is due to the migration constraint in Equation 1 being violated.

We also measured the budget utilization of the migration thread while it was active. Results are shown in

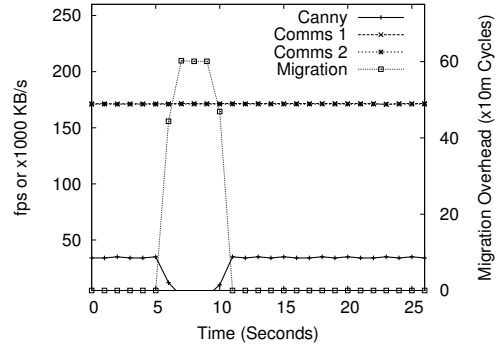

Fig. 10. Migration With Added Overhead

| Variables | $E_s$ | $\Delta_s, worst$ | $\Delta_s, actual$ | $C_m$ | $T_m$ |
|---|---|---|---|---|---|
| Time (ms) | 79.8 | 891.4 | 825.1 | 10 | 50 |

TABLE III
MIGRATION CONDITION WITH ADDED OVERHEAD

Table IV for the interval [6s,10s] of Figure 10. As can be seen, the migration thread budget consumption peaked at 91.5%. Scheduling and accounting overheads prevent the migration thread from actively using its entire budget. We are currently working on optimizations to our system to reduce these overheads.

| Time (sec) | 6 | 7 | 8 | 9 | 10 |
|---|---|---|---|---|---|
| Utilization | 67.5% | 91.5% | 91.5% | 91.5% | 71.5% |

TABLE IV
MIGRATION THREAD BUDGET UTILIZATION

The same experiment was repeated without a dedicated migration thread. Migration was instead handled in the context of an IPI handler that runs with interrupts subsequently disabled. Consequently, the handler delays all other threads and their VCPUs during its execution, as shown in Figure 11.

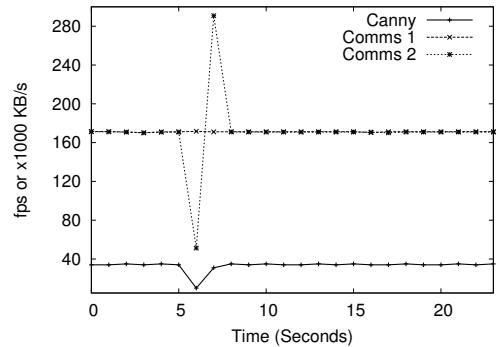

Fig. 11. Migration Without a Migration Thread

Finally, Table V shows the values of the variables used in Equation 1 when the migration overhead first starts to impact the Canny frame rate. In theory, the minimum $\Delta_s$ that violates Equation 1 in this case is 20ms. However, because $\Delta_s, worst$ is a worst-case estimation, the first visible frame rate drop happened when it reached 26.4ms. At this time, the actual (measured) budget consumption of the migration thread is 19.2ms, which is smaller than 20ms because of the scheduling and accounting overheads mentioned earlier.

| Variables | $E_s$ | $\Delta_s, worst$ | $\Delta_s, actual$ | $C_m$ | $T_m$ |
|---|---|---|---|---|---|
| Time (ms) | 79.8 | 26.4 | 19.2 | 10 | 50 |

TABLE V
MIGRATION BOUNDARY CASE CONDITION

### B. Predictable Communication

Here, we constructed 5 different scenarios according to Figure 6 and tried to predict the worst case round trip time using Equation 3. The VCPU settings of the sender and receiver, spanning two different sandboxes, are shown in Table VI.

| Case # | Sender VCPU | Receiver VCPU |
|---|---|---|
| Case 1 | 20/100 | 2/10 |
| Case 2 | 20/100 | 20/100 |
| Case 3 | 20/100 | 20/130 |
| Case 4 | 20/100 | 20/200 |
| Case 5 | 20/100 | 20/230 |

TABLE VI
VCPU PARAMETERS

In addition to the VCPU parameters, we also calculated the values of $M$, $N$, $\delta_s$ and $\delta_d$ by setting the message size to 2KB for both sender and receiver (i.e. $M = N = $ 2KB) and *disabling caching* on the test platform. The message processing time $K$ has essentially been ignored because the receiver immediately sends the response after receiving the message from the sender.

Both sender and receiver threads running on VCPUs $V_s$ and $V_d$, respectively, sleep for controlled durations, to influence phase shifts between their periods $T_s$ and $T_d$. Similarly, the sender thread adds busy-wait delays before transmission, to affect the starting point of communication within its VCPU's available budget, $C_s$. After 10000 message exchanges in each scenario, we recorded the observed worst case round trip time shown in Figure 12. As can be seen from Figure 12, the observed worst case time recorded in Quest-V is within the bounds, but also close, to the prediction derived from Equation 3.

## VI. RELATED WORK

Quest-V is a multikernel designed to work on modern multicore processors. It has similarities to Barrelfish [1], Factored OS (FOS) [12], Hive [13] and Corey[3]. Corey is a library OS that allows processing cores to be dedicated to applications, which can communicate via shared memory. Hive provides fault containment within groups of processing nodes, or *cells*. These systems all address scalability by eliminating as much as possible the sharing of system state across cores. Quest-V differs from these systems by using virtualization for fault isolation, and methods to enforce real-time task execution.

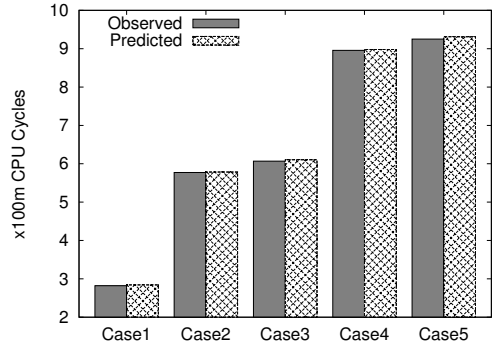

Fig. 12. Worst Case Communication Time Prediction

The seL4 microkernel [14], [15] attempts to statically verify software faults will never occur. This contrasts with Quest-V, which uses sandboxing techniques to isolate the effects of faults at runtime. As of yet, we are unaware of seL4 for multicore processors. While efforts have been made to derive WCET bounds for interrupt-response latency in seL4, the work in this paper is concerned with establishing bounds on communication and migration.

In Quest-V, each sandbox kernel performs its own localized scheduling, without requiring the overhead of hypervisor intervention. This contrasts with virtual machine monitors such as Xen [2] that schedule and manage the assignment of guest VCPUs on PCPUs. Quest-V adopts a *semi-partitioned* strategy [16], [17], allowing task migration to remote sandboxes. Quest-V's migration scheme is intended to maintain predictability, even for tasks that may have started executing on VCPUs in one sandbox and then resume execution in another. Other systems that have supported migration include MOSIX [18] and Condor [19], but these do not focus on real-time migration.

In other work, reservation-based scheduling has been applied to client/server interactions involving RPC [20]. This approach is based on analysis of groups of tasks interacting through shared resources accessed with mutual exclusion [21], [22]. A bandwidth inheritance protocol is used to guarantee the server handles each request according to scheduling requirements of the client. We intend to investigate the use of bandwidth inheritance protocols across sandboxes, although this is complicated by the lack of global prioritization of VCPUs. In this work, we have focused instead on deriving a delay bound for round-trip communication, in the absence of a global scheduler or system-wide clock.

## VII. CONCLUSIONS AND FUTURE WORK

This paper describes the mechanisms in Quest-V to support real-time thread migration and communication. The work builds on ideas first considered in our OSPERT paper [23] but which have now been implemented. The "distributed system on a chip" design of Quest-V ensures separate sandbox kernels can coexist and remain functional in the presence of faults in other sandboxes. Hardware virtualization is used to isolate sandboxes, which perform local I/O management and scheduling.

Quest-V allows threads to migrate between sandboxes, either due to performance, predictability, resource affinity or fault recovery reasons. We have shown how Quest-V's migration mechanism between separate sandboxes is able to ensure predictable VCPU and thread execution. Experiments show the ability of our Canny edge detector to maintain its frame processing rate while migrating from one sandbox to another. This application bears significance in our RacerX autonomous vehicle project that uses cameras to perform real-time lane detection.

Finally, we have shown how Quest-V is able to enforce predictable time bounds on the exchange of information between threads mapped to different sandboxes. This lays the foundations for real-time communication in a distributed embedded system. Future work will investigate *lazy* migration of only the working set (or *hot*) pages of address spaces. This will likely reduce initial migration costs but incur page faults that will need to be addressed predictably.